\documentstyle[aps]{revtex}
\include{psfig}

\begin{document}
\draft
\preprint{HEP/123-qed}
\title{ A soluble model of evolution and extinction  dynamics
in a rugged fitness landscape.
}
\author{Paolo Sibani\cite{byline}\\
 }
\address{Dept. of Mathematical Sciences\\
San Diego State University CA, USA 
}
 
\date{\today}
\maketitle
\begin{abstract}
We consider  a continuum  version of  a previously
introduced and numerically studied     model 
of macroevolution \cite{Sibani95},
in which agents evolve by an optimization process in a 
rugged fitness landscape and die due to  
their competitive interactions. We first formulate
dynamical equations for the fitness distribution and the
survival probability.
Secondly we   
  analytically   derive
  the $t^{-2}$ law which characterizes
the life time   distribution of
 biological  genera. Thirdly  we 
 discuss  other  dynamical properties
 of the model  as the rate of extinction and
 conclude with a brief discussion. 
\end{abstract}

Aspects of evolution and extinction can be described 
as emergent behavior in a large set of interacting 
agents\cite{Sibani95,Bak93,Sneppen95,Newman95}, moving 
stochastically in a rugged  
  fitness landscape\cite{Wright82}.  
The behavior of  the models  of
Refs.\cite{Bak93,Sneppen95,Newman95} 
 stems from fluctuations in a time 
homogeneous stochastic 
process. This   agrees with a commonly held 
perception, e.g. implied when a
 birth-death process with constant rates\cite{Raup78}
is used to fit  survivorship data and when the
size of extinction events is presented as a   `kill curve'\cite{Raup95}. 
A quite different paradigm  is  also 
frequently met in the literature:
Raup and Sepkoski \cite{Raup82} noted that
 the apparent decrease of the  extinction rate
through geological times could be 
`...  predictable from first principles if one argues that
general optimization of fitness through evolutionary time
should lead to prolonged survival '.  Gould\cite{Gould96}
uses an unexpected source of statistical data
to illustrate evolutionary non-homogeneity
as it reveals itself in the 
`unreversed, but constantly slowing, improvement
in mean fielding average through the history of baseball'.
Concurring  observations from 
 experimental studies of bacterial
 evolution  in a constant environment can be
 found in  Ref.\cite{Lenski95} as well as  from  numerical
 experiments on the `long jump dynamics' of the  NK model
 in Ref.\cite{Kauffman95}.

In this Letter we consider stochastic evolution in a rugged 
fitness landscape. The 
  assumptions are the same  in spirit
 as those  of a previously introduced
and numerically studied  `reset'  model\cite{Sibani95}.
However, they are  here expressed    in a
further simplified way, allowing  
a   (mainly) analytical rather than (mainly) numerical
treatment, and leading to close form expressions
for the survivorship curves and life span  distributions,
which are of  general interest in the study 
of complex evolving systems, biological or not. 
We use two -- somewhat  extreme  -- assumptions in line
with a non-stationary evolution paradigm: Firstly, the progeny
of individual mutants less fit than the currently
dominating  genotype  {\em never} establish itself
within the population. Then, as   
a macroscopic evolutionary step can only be  triggered 
by a fitness  {\em record}  within the population,   
the current typical genotype always  codes 
the  best  solution found `so far'. 
 Secondly, competitive interactions among
 species depend on fitness 
in a  non-symmetric way, as evolving  species  
only kill their less fit neighbors.
The  predictions of the present model
resemble the behavior of the reset model and are in good
 agreement with empirical data describing
biological genera
\cite{Raup78,Raup95,Stanley78,Raup85,Raup86,Baumiller93}.

In the sequel  we first derive    equations for
the fitness distribution of  the system $P(x,t)$  
and  for  the  probability $W_t(\tau)$
that  a tagged species born at time $t$  survive time $\tau$.
We then analytically find the $\tau^{-1}$ dependence of $W$
and   the ensuing  $\tau^{-2}$ dependence
of the life-time distribution $R_t(\tau)$.
Next we discuss the   parametric $t$ dependence
which is not in general 
analytically available,   the
effect of averaging over $t$, and the long time
asymptotic behavior of $P(x,t)$  for different
parameter values. 
We conclude by with a brief assessment of the
robustness of  the model.

To construct  a dynamical equation 
for    $P(x,t)$  
we proceed  in two steps, starting
with the  limiting  case where  no extinctions take place
and where, as a consequence of
hill climbing in a random fitness landscape,
a suitably defined\cite{Sibani95,Sibani93,Kauffman87}
  average  fitness 
grows logarithmically:
\begin{equation}
D(t) = \log(t+1).
\label{deterministic}
\end{equation}

With no interactions, 
an  initial fitness distribution  would  be
rigidly  shifted  in (log) time.
As $D$ solves
the   equation of motion 
$v(x) = d x/d t = \exp(-x)$
  with initial condition $D(0) = 0$, the
time evolution  of a  distribution of
non interacting agents $ P(x,t)$   solves   
the  transport  equation:
$\partial  P(x,t) / \partial t + \partial(v(x) P(x,t)) / \partial x = 0.$
Interactions enter via   an additional    term 
$ - g  P(x,t) K(P(x,t))$, where $K$
is an effective killing rate and where 
the  constant  $g $ 
describes what  fraction of the system is affected by 
an evolutionary event.

Species going  extinct vacate a niche, which is refilled 
at a later time.
This  in and outflow is expediently accounted by introducing
 a   `limbo' state, which absorbs extinct species, 
and from which new species emerge at 
 the  low fitness boundary  of the  system. 
A finite upper bound to   the total number of species  
which  can coexist implies a 
conservation law :  $ N(t) + \int_0^\infty P(y,t)dy = 1 $.
 With the chosen normalization 
  $N(t)$ is   the fraction of species in the  limbo  state, while
$P(x,t)$ is the probability density of finding a living species with fitness 
$x$.
The above  considerations lead us to   
the differential   equations:
\begin{eqnarray}
\frac{ dN(t) }{dt} = - b  N(t) + g  \int_0^\infty P(z,t) K(P) dz 
\label{PDE0}\\
\frac{ \partial P(x,t)}{\partial t} = -\frac{\partial (v(x) P(x,t))}{\partial x} - 
g  P(x,t)  K(P)
\label{PDE}
\end{eqnarray}
where $ b $ is the rate at which species are generated at the low fitness end 
of the system.
The  corresponding initial and boundary  conditions are:
$N(t=0) = N_0, \;
\forall x: P(x,t=0) = P_0(x), \; 
\forall t: \int_0^\infty P(x,t) dx = 1 - N(t) < \infty, \;$
and finally 
$\forall t: P(x=0,t) =  b  N(t)$.

We consider below a form of the killing rate $K$
which is as close as possible
to the reset model: The killing at fitness $x$ is taken to depend
on the rate of evolutionary change of agents with fitness 
larger than $x$: low-fitness agents suffer
if  high fitness agents evolve - but not vice versa.
This leads to
\begin{equation}
K(v(x)P(x,t)) = (-\int_x^\infty \partial(v P)/\partial x)^\alpha  dx = (v(x)P(x,t))^\alpha,
\label{kill} 
\end{equation}
simply expressing the killing rate 
as the evolutionary current raised to a power.
The exponent  $\alpha$ just  introduced  
allows more  generality without unduly complicating the analysis:
It  accounts in a simplified way for 
 possible (spatial) correlations   effects  in a model where
information about individual species is retained.
If $\alpha <1$ $ (>1)$, a move by an old,  slowly evolving  species 
triggers  a larger (smaller ) cascade of extinctions  than 
one by a  young, fast evolving species. 
Figure 1 shows six snaphots of the fitness distribution
resulting from the above equations,  
at times equally spaced on a logarithmic scale and  for $\alpha =1$,
$b=1$ and $g=40$.   

A quantity often used to characterize  paleontological 
data is the survivorship  curve of a cohort or the closely related
life span distribution\cite{Raup78}.
In our treatment the former quantity  corresponds to 
the probability $W_t(\tau)$ that an agent 
appearing at time $t$ survive time $\tau$, while
the latter can be  found
from $W_t(\tau)$ by differentiation:
\begin{equation}
R_t(\tau ) = -\frac{d W_t(\tau)}{d\tau }.
\label{LIFESPAN}
\end{equation}

As an agent born at $t$ and  alive at  time $t + \tau$ 
invariably has  fitness $D(\tau) = \ln (\tau +1) $ and as
 the  probability of being killed in the interval $d\tau$ 
is $K( P(D(\tau),t+\tau) ) d\tau$,
$W$ must obey the differential equation:
\begin{equation}
\frac{d \ln W_t(\tau)}{d\tau } = - g  K( P(D(\tau),t+\tau) )\; \; \tau >0,
\label{survival_prob}
\end{equation}
with initial condition $W_t(\tau = 0)=1$.

Finally, the model extinction rate is simply the
 fraction  of species which die per unit of time,  at time $t$:
\begin{equation}
r(t)  = g  \int_0^\infty P(x,t) K(P(x,t)) dx = dN/dt +  b  N.
\label{ext_rate}
\end{equation}
Note that if  $ b   \rightarrow \infty$, then
 extinct species are immediately  replaced, as in
 Ref.\cite{Sibani95}. Furthermore for any $b$ and
 large $bt$ $dN/dt$ is negligible and   
$ b  N(t) \rightarrow r(t)$ so that the extinction  
closely balances the inflow.

As a first step towards the solution of Eq.\ref{PDE},
we set  $q = v P$ and notice  that
$q$ can be written as 
 $q(z(x,t))$ with
\begin{equation}
dq/dz = -g  q^{\alpha +1},
\label{q_eq}
\end{equation}
and where $z(x,t)$ satisfies  
\begin{equation}
\partial z/\partial t + v(x) \partial z / \partial x = 1.
\label{z_eq}
\end{equation}
The  solution of Eq.\ref{q_eq} is simply $q = (\alpha g  z )^{-1/\alpha}$.  
To solve Eq.\ref{z_eq}  we let  $A$ and $B$ be any two  
functions of a single real variable $x$, which
are   continuous for $x>0$ and which  vanish identically  for  $x<0$.
For $v = \exp(-x)$, the general solution has the form 
$z(x,t) = \epsilon \exp(x) + (1-\epsilon ) t + A(t+1-\exp(x))+ B(\exp(x) -(t+1))$
for some constant $\epsilon < 1$.
Utilizing the  initial and boundary conditions, we  find
$
A(y) = ( b  N(y))^{-\alpha} , \; \; y>0
$
and 
$
B(y) = (y+1)^\alpha P_0( \log(y+1))^{-\alpha} - g  \alpha y,\; \; y>0, 
$
leading to 
\begin{equation}
P(x,t) = \frac{e^x}
{\left[
 g  \alpha t +(e^x-t)^\alpha P_0^{-\alpha}(\log(e^x -t))
 \right] ^{1/\alpha}} 
\label{sol2a} 
\end{equation} 
for $ x > D(t)$,  while for $ x < D(t)$ we have 
\begin{equation}
P(x,t) = \frac{e^x}
{\left[
 g  \alpha (e^x-1) + ( b  N(t+1 - e^x))^{-\alpha}
 \right] ^{1/\alpha}} 
\label{sol2b}
\end{equation}
Note that  $P$ is continuous in $x$, although its derivative will
in general be discontinuous at $x = D(t)$.

The survival probability of a species born at time $t$
(the survivorship curve of a cohort\cite{Raup78}), can be  
obtained analytically by solving Eq.\ref{survival_prob}.
This is so because when inserting
$D(\tau)$ in lieau of $x$ in Eq.\ref{sol2b}, the $\tau $ 
dependence  in the argument of the (unknown) function $N$ drops out.
The solution is:
\begin{equation}
W_t (\tau ) = \left[ 
\frac{( b  N(t))^{-\alpha} }{( b  N(t))^{-\alpha}+g  \alpha \tau } 
\right]^{1/\alpha}.
\label{survive} 
\end{equation}
As  $W$  vanishes for large $\tau$,  all  species eventually die,
regardless of the value of $\alpha$.
This behavior 
is very desirable from a modeling point of view, as it agrees 
with the fact that  by far the largest
number of species which ever lived are now extinct \cite{Raup86}.
The    distribution of life spans can be obtained from
Eq.\ref{survive}   by differentiation, as expressed in Eq.\ref{LIFESPAN}. 
 If $\alpha $ is close to unity,
we find a $\tau^{-1}$ behavior  for $W_t(\tau)$, and hence 
a $\tau^{-2}$ for $R_t(\tau)$, independently of $t$. 

Averaging these distributions with respect to 
$t$ over a time window $T$ is needed
if  the time of appearance  of  species 
is not precisely known, or if data are scarce.
Weighing 
$R_t(\tau )$  by  the normalized rate at which
new species flow into the system we obtain:
\begin{equation}
R(\tau) = \frac{\int_0^{T-\tau} N(t) R_t(\tau) dt}{\int_0^T N(t) dt}. 
\label{lifetime1}
\end{equation}
Of course,    averaging 
 does not change the behavior 
   significantly if 
$T$ is  short compared
to the typical lifetime of the species.
In the opposite limit, 
the behavior is also maintained
if  $N(t)$ does not  not vanish 'too rapidly'
in the limit  $t \rightarrow \infty$.
To better appreciate  this last point, we use  Eq.\ref{lifetime1} 
in conjunction with Eqs.\ref{survive}æ
and   \ref{LIFESPAN}, and  express 
$N^{-\alpha}$ by $W_t(\tau)$, obtaining:
\begin{equation}
R(\tau) = g  (g  \alpha \tau)^{-1-1/\alpha}
\frac{
  \int_0^{T-\tau }
 (1 - W_y^\alpha(\tau))^{1+1/\alpha} dy
}
{\int_0^{T-\tau } b  N(y) dy }
\label{mellem}
\end{equation}

Even though this integral cannot be evaluated explicitly,  
 Eq.\ref{survive}
shows that the 
$\tau$ dependence of the integrand  
is negligible
if the inequality $\alpha g  \tau > ( b  N(y))^{-\alpha} $
holds  throughout the integration interval.
The $\tau $ dependence stemming from  the limits of the
integrals can also be ignored
for  $\tau << T$. Hence
$R \propto \tau^{-1 - 1/\alpha}$,  
similarly to  the non-averaged case.
As shown later, 
when  $\alpha $ close to unity and $g$
sufficiently large,  the model yields
$r(t) \approx   b  N(t) \geq t^{-\delta}$ ,
with $\delta$ close to $0.5$, which means
that even though  $\tau <<T$ the relation
 $ \tau > T^{\delta \alpha}$ can be fulfilled.

We now restrict ourselves 
to  a limiting  case in which
$N(t=0 ) = 1$  which is formally at variance
with our boundary conditions. However, a limiting 
process shows that 
the relevant expression for $P(x,t)$ for 
 $x < D(t)$ remains  Eq.\ref{sol2b}, while $P = 0$ 
for $x >D(t)$.
A non-linear  equation for   $N(t)$ 
is now 
obtained by integration of    Eq.\ref{sol2b}, followed  by 
a change of variables.
The result is
\begin{equation}
1 - N(t) = \int_0^t \frac{ dy}{
\left[g  \alpha (t-y) + ( b  N(y ))^{-\alpha} \right]^{1/\alpha} 
}\label{eqforN}\\
\end{equation} 
Differentiating Eq.\ref{eqforN} with respect to time, and utilizing 
Eq.\ref{ext_rate}, we find 
the extinction rate: 
\begin{equation}
r(t) = \int_0^t \frac{g  \; dy}{
\left[g  \alpha (t-y) + ( b  N(y))^{-\alpha} \right]^{1 +1/\alpha } 
} \label{Eqforr}\\
\end{equation}

A closed form 
solution of these (equivalent) integral equations
could not be found in the general case. We notice  
however  a major difference
in the asymptotic behavior for
$\alpha <1$ and $\alpha \geq 1 $.
In both cases the time independent  function $P_\infty(x)$ 
obtained  by taking 
$t \rightarrow \infty$  and by setting $N(t) = N_\infty \neq 0$ in 
Eq.\ref{sol2b} formally satisfies the model equations.  
However, only for  $\alpha < 1$ is  $P_\infty(x)$ normalizable and
thus  a true solution. The corresponding steady state value of  $N$, 
  $N_\infty$  is then implicitly  given by the relation
$1-N_\infty = ( b  N_\infty)^{1-\alpha}/(g (1-\alpha))$, which
always has a solution in the unit interval. 

In the case 
$\alpha \geq 1$,  normalizability of $P(x,t)$ 
requires that   $N(t) \rightarrow 0 $ for $t \rightarrow \infty$.
No steady state solution can then exists, since 
$P(x,t)$ vanishes with $t$ at any fixed $x$,
as e.g. in the familiar case of simple diffusion on the infinite line.
For $\alpha <1$,  the  steady state solution is strictly speaking  only
approached logarithmically due to the form of $D(t)$. Neglecting this 
logarithmic corrections  we see from Fig. 2 a power-law approach to
a quasistationary behavior over a substantial time range. 

For long times 
$\alpha = 1$ and  $ b  t >> 1$   the term
$dN/dt$ in Eq.\ref{PDE0} is negligible, and 
 $ b  N(t) \approx r(t)$.
In this limit  we can  also  neglect $N$ compared to one, thus  
finding the following  approximate equation for  $r(t)$:
$
1 = \int_0^t \frac{g  dy}{g  (t-y) + r(y)^{-1}}
$
which  has the solution  $r(t) = (g  t)^{-1}$.

Fig. 2 shows a  $r(t)$ vs. t for
$\alpha = 0.95$ $b = 1$ and several $g$ values.
As noted, for a wide time span,  
$r \propto t^{-\gamma}$
where  $\gamma$ decreases with increasing 
$g$,
similarly  to the result obtained in  the
the  simulations of the reset model\cite{Sibani95}.
No qualitative changes are observed when varying
$\alpha$ in a small range below one, or when  changing  $b$. 
In summary, for $\alpha$ slightly below one, 
and $g$  is sufficiently large, 
the life span distribution (averaged or not) decays algebraically  
with an exponent slightly above 
$-2$ and  the rate of extinction decays with an exponent close 
to $-0.5$ untill it reaches a regime of hardly detectable change. 

The most comprehensive empirical life span distribution
available, comprising  about 17500 extinct genera
of marine animals has been tabulated by Raup\cite{Raup95}
from data compiled by   Sepkoski\cite{Sepkoski89}.
 These data cover  about 100 million years and  
 display a  very clear $t^{-2}$ dependence in a log-log
plot \cite{Sibani95,Sneppen95} over this range, which 
concurs with the behavior of our $R_t(\tau)$.
More recent analysis  by Baumiller\cite{Baumiller93} 
of several data set describing
crinoid survivorship - our $W_t(\tau)$ - over a comparable
time span  in part concur  with a $t^{-1}$ law,  and hence
with a  $t^{-2}$  law  for the life-time distribution.
 Finally, 
survivorship curves for european mammals were considered by 
Stanley \cite{Stanley78}. These data span approximately 3 million years
stretching to the W\"{u}rm period and include much fewer species.
 The distribution of lifetimes deviates 
from a $t^{-2}$ law  by having  an extra 'hump' 
approximately in the middle
of the time range.
 
Paleontological data are  commonly interpreted  
using a  birth and death
model\cite{Raup78,Baumiller93}, in which 
non-interacting
species are born and die
  with two distinct    constant rates 
of speciation and extinction, 
$\lambda$ and $\mu$, and where 
the genus  becomes  extinct
together with its  last species.
Interestingly,  the survivorship formula 
generated by this  model is,  
 for   $\lambda = \mu$ and for an initial number
of species equal to one,  
 identical
to  our Eq.\ref{survive} - with
$\alpha = 1$,
{\em  as far as its  dependence on the life time} -our $\tau$- goes.
By continuity 
so are the model predictions  in the often recurring
situation when  $\lambda \approx \mu$. The similarity 
in the formulae is however contingent to the initial condition 
 and should be regarded as accidental\cite{note1}.
 
In line with the conclusion  of the
reset model, we have shown analytically  that 
a large body of data describing evolution
on  coarse scales of time and taxonomical level
can be explained by  two very simple ideas: 1) that 
fitness records in  random searching trigger
evolutionary events, and
 2) that  the   species competition is `asymmetric',
 with high fitness species being more resilient.
 
The robustness of this approach has  already been analyzed 
to some extent: The effect of    additional
and externally imposed random  killings of a 
fraction of the  agents -- mimicking  catastrophies --
has been studied by M. Brandt\cite{Brandt97}, who found that  
 the   life-span distribution was  not 
 affected. This is to be  expected,  as even very large
mass extinction events  - in the model  as well as 
in reality - only account for  a small fraction of all extinctions.
We also  explored    other choices 
of the killing term in Eq.\ref{kill}, finding  that 
the $1/t^2$ law disappears if   the  asymmetry of the 
interspecies interactions
is removed,   with the possible exception
of special   values  of the  coupling constants.

After this paper was submitted the author  became aware of a preprint
by Manrubia and Paczuski\cite{Manrubia97}, which also  
treats  evolution and extinctions by means of a transport equation,
an  finds a $t^{-2}$ life-time distribution,  albeit the basic  
dynamical mechanism is quite different from ours. 
 
\noindent{\bf Acknowledgments}\\                
I would  like to thank Preben Alstr\o m, Michael Brandt,
Peter Salamon and Jim Nulton for useful
conversations. This work was supported by the 
Statens  Naturvidenskabelige Forskningsr\aa d.

\newpage
 
\begin{figure}[t]
\centerline{\psfig{figure=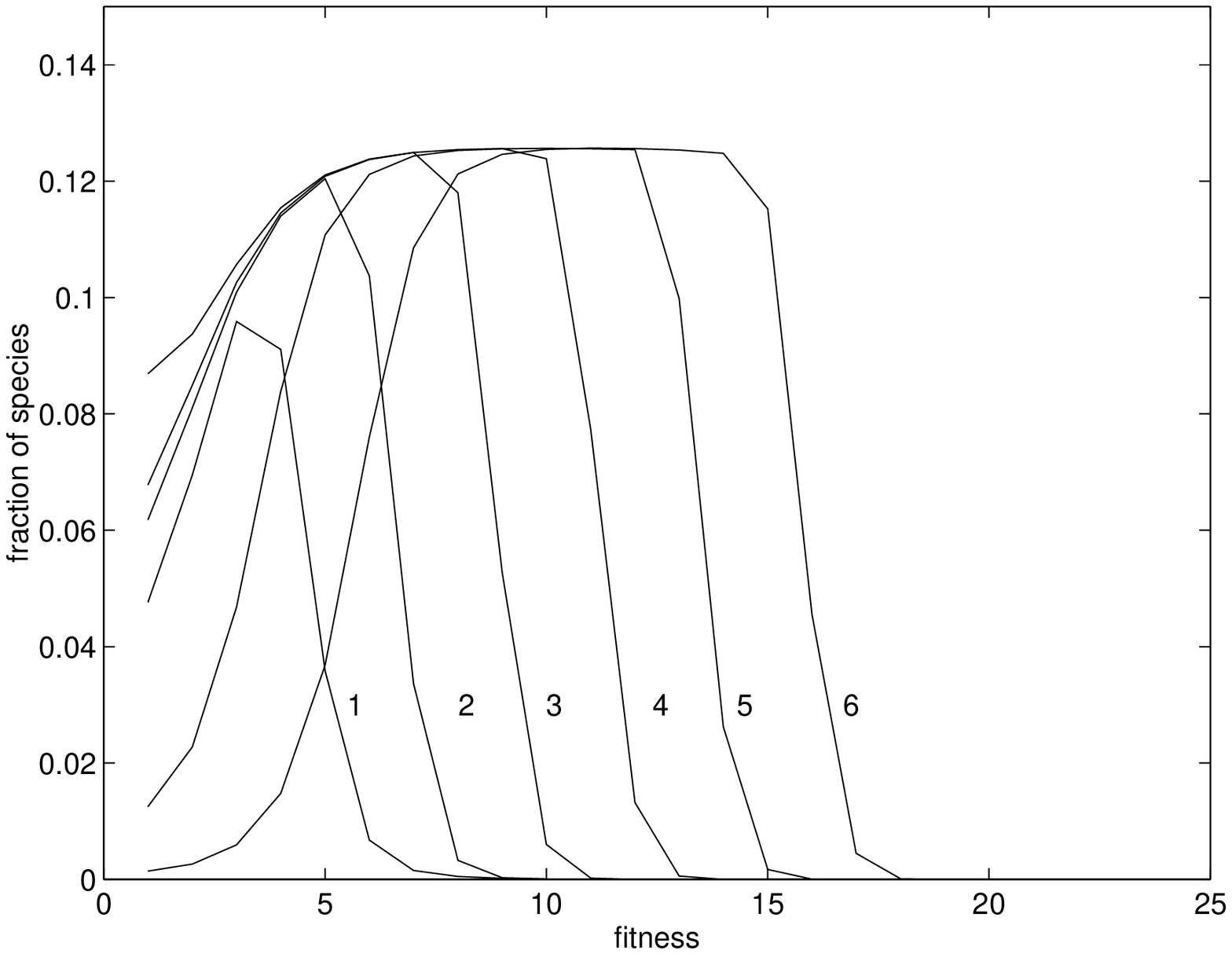,height=8cm,width=9cm}}
\caption{
 For $1 \leq n \leq 6$ the $n$'th plot   in  the figure
 depicts the fitness distribution $P(x,t)$  at time $t=10^n$ 
  The parameters used are $\alpha =1$, $b=1$ and $g=40$.}  
\end{figure}

\begin{figure}
\centerline{\psfig{figure=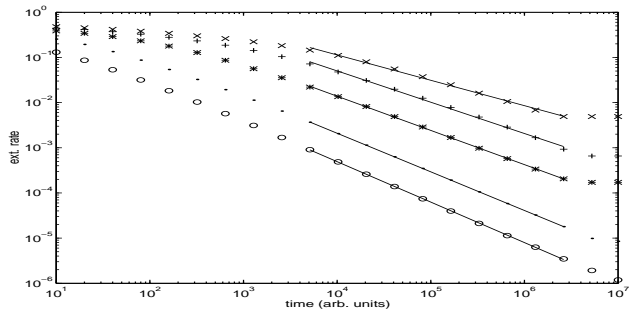,height=8cm,width=9cm}}
\caption{
The numerically obtained extinction rate is plotted vs  time,
for $ b  = 1$,  $\alpha = 0.95$ and for  $g=5$ ($o$),
$10$ ($.$), $20$ ($*$), $30$ ($+$)  and
$40$ ($\times$).
The decay is approximately a power-law with exponents  
equal to $-0.89, -0.85, -0.75, -0.69 $ and $ -0.56$. 
These slopes are visualized  by
the full lines.} 
\label{Fig.2}
\end{figure}

\begin{references}
\bibitem{note1}
The 'birth and death' formula \cite{Bailey64} 
depends exponentially 
on the  number $a$ of species initially in the genus.
In the application one needs $a=1$.
It is also  appreciated 
  that this  model cannot  account 
for the appearance of  new genera, as the 
parameter choice $\mu > \lambda$  used  to
fit the data would otherwise imply the total
 extinction of life\cite{Raup78,Baumiller93}.
\end{references}
\end{document}